\documentclass[twocolumn]{autart}%\documentclass[11pt]{article}

\usepackage{times}
\usepackage{graphicx}        % standard LaTeX graphics tool
\usepackage{amssymb,amsmath,mathrsfs}
\usepackage{mathtools}
\usepackage{enumerate}
\usepackage{enumitem}
\usepackage{framed}
\usepackage{algorithm,algorithmic,psfrag}
\usepackage{soul}
\usepackage{color}
\usepackage{wrapfig}
\usepackage{arydshln}

\newtheorem{assumption}{Assumption}
\newtheorem{lemma}{Lemma}
\newtheorem{theorem}{Theorem}

\definecolor{paleGreen}{rgb}{.3, .7, .3}
\definecolor{coolBlue}{rgb}{.3, .5, 1}
\definecolor{rosePink}{rgb}{.9, .5, .4}

\newcommand{\cov}{\mbox{\rm cov}}
\newcommand{\sise}{\mbox{\textnormal{\tiny SISE}}}

\newif\ifsolutions
\solutionsfalse

\begin{document}
\begin{frontmatter}
\title{Simultaneous input \&\ state estimation, singular filtering and stability\thanksref{footnoteinfo}} % Title, preferably not more 
                                                % than 10 words.

\thanks[footnoteinfo]{The authors would like to thank for financial support from CINELDI -- the Centre for Intelligent Electricity Distribution, Research Council of Norway project no. 257626/E20. Author~2 thanks Kuwait University for funding. This paper was not presented at any IFAC meeting. Corresponding author R.R. Bitmead. Tel. +1 858 822 3477.}
%,
%Fax +1 858 822 3107.}

\author[NTNU]{Mohammad Ali Abooshahab}\ead{mohammad.ali.abooshahab@ntnu.no}
\author[UCSD]{Mohammed M.J. Alyaseen}\ead{malyasee@eng.ucsd.edu}
\author[UCSD]{Robert R. Bitmead}\ead{rbitmead@ucsd.edu}
\author[NTNU]{Morten Hovd}\ead{morten.hovd@ntnu.no}

\address[NTNU]{Department of Engineering Cybernetics, Norwegian University of Science \&\ Technology, 7491 Trondheim, Norway.}
\address[UCSD]{Department of Mechanical \&\ Aerospace Engineering, University of California, San Diego, 9500 Gilman Drive, MS-0411, La Jolla CA 92093-0411, USA.}

                             % keyword list or with the help of the Automatica keyword wizard

\begin{abstract}
Input estimation is a signal processing technique associated with deconvolution of measured signals after filtering through a known dynamic system. Kitanidis and others extended this to the simultaneous estimation of the input signal and the state of the intervening system. This is normally posed as a special least-squares estimation problem with unbiasedness. The approach has application in signal analysis and in control. Despite the connection to optimal estimation, the standard algorithms are not necessarily stable, leading to a number of recent papers which present sufficient conditions for stability. In this paper we complete these stability results in two ways in the time-invariant case: for the square case, where the number of measurements equals the number of unknown inputs, we establish exactly the location of the algorithm poles; for the non-square case, we show that the best sufficient conditions are also necessary. We then draw on our previous results interpreting these algorithms, when stable, as singular Kalman filters to advocate a direct, guaranteed stable implementation via Kalman filtering. This has the advantage of clarity and flexibility in addition to stability. En route, we decipher the existing algorithms in terms of system inversion and successive singular filtering. The stability results are extended to the time-varying case directly to recover the earlier sufficient conditions for stability via the Riccati difference equation.
\end{abstract}
\end{frontmatter}
\maketitle

\section{Introduction}Simultaneous Input and State Estimation (SISE) algorithms take a system with an unknown disturbance input sequence, $\{d_t\}$, and measured output signal, $\{y_t\}$, plus possibly also a known input sequence, $\{u_t\}$, and produce estimates of both $d_t$ and the system state $x_t$, based on fixed lag smoothing. There have been a great number of recent papers on these algorithms, with  a recent subset \cite{MarroZattoniAutom2010,FangShiYiIJACSP2011,FangDeCallafon:Autom2012,YongZhuFrazzoliAutom2016} focusing on conditions for stability. A feature of SISE algorithms is that the disturbance signals have uncertain provenance. So SISE presumes that no model is available for this signal and the algorithm proceeds without an explicit description of the disturbance signals' statistical properties. Thus techniques such as extended state observers \cite{BZGuoZLZhaoBook2016} and augmented Kalman filters \cite{Anderson&Moore:79,Simon:2006} are inapplicable, as both rely on disturbance models. A recent paper by the authors \cite{BitmeadHovdAbooshahab:Autom2019} establishes that, when stable, the linear system SISE algorithm of \cite{GillijnsDeMoorAutom2007a} coincides with the Kalman filter with $\{d_t\}$ modeled by white Gaussian noise of unbounded variance. Various approaches consider first estimating the state \cite{MarroZattoniAutom2010} and then using the state recursion to reconstruct $d_t$, or estimating the disturbance first and then reconstructing the state \cite{GillijnsDeMoorAutom2007a,GillijnsDeMoorAutom2007b,YongZhuFrazzoliAutom2016}. These methods rely on geometric approaches and system inversion, although there is a strong overlap with least-squares state estimation concepts of unbiasedness and optimality.

SISE algorithms go back to least to Kitanidis \cite{KitanidisAutom1987} with antecedents \cite{JDGloverTAC1969,MendelTAC1977,SanyalShenTAC1974} concentrating on input signal reconstruction. Here, we follow the formulation from Yong, Zhu and Frazzoli \cite{YongZhuFrazzoliAutom2016}, which in turn builds on \cite{GillijnsDeMoorAutom2007a,GillijnsDeMoorAutom2007b}. We consider linear time-invariant systems to add clarity and to explore the connection to optimal estimation before extending to uniformly time-varying systems.

Input estimation is a signal processing technique associated with deconvolution of measured signals after filtering through a known dynamic system. Examples include the estimation of rainfall given river flow and the calculation of salinity in the ocean accommodating for sensor dynamics \cite{FangDeCallafonCortesAutom2013}. Here, the central objective is to estimate the driving disturbance signal $d_t$ and there is little interest in the sensor state. The algorithm should be stable, however. Our particular driving problem, on the other hand, is the estimation of generator states in part of a power grid when the interconnection signals are unknown \cite{AliMortenBobCCTA2019}. Here the priority is to estimate network generator states in the face of unmodeled and unmeasured consumption, which is treated as the disturbance signal. In spite of these distinct objectives, the same algorithms have been used.  

\subsection*{Contributions \&\ organization}
Our objective in this paper is to attempt to bring some clarity and unity to this picture by establishing precisely the connection to system inversion and optimal estimation by deriving necessary and sufficient conditions for stability using explicit system inverse formul\ae\ and algebraic Riccati equations, starting with the time-invariant case. Earlier stability conditions were sufficient only but derived in the time-varying situation. We recover these. Further, when stability is not achieved by these SISE algorithms, we propose a modification based on inner-outer factorization, which maintains state estimation performance at the expense of simple disturbance recovery. This can be compared with the techniques advanced in \cite{MarroZattoniAutom2010} for approximate system inversion with delay. Beyond this work, we know of no other which addresses estimation when the stability conditions fail.

Section~\ref{sec:problem} presents the SISE problem for a linear time-invariant system. Section~\ref{sec:zdf} studies the zero direct feedthrough case and the corresponding SISE of \cite{GillijnsDeMoorAutom2007a} and shows that, in the square case where the number of measurements equals the number of disturbance channels, the input estimator is the inverse of the $d_t$-to-$y_t$ system and the state estimator is a plant simulation. Stability depends on the transmission zeros of the former system. These necessary and sufficient stability conditions then are extended to the non-square case with more measurements. This involves the Riccati difference equation and a detectability condition. Section~\ref{sec:ndf} expands this analysis to the full-rank direct feedthrough case and comments on the non-full-rank case of \cite{YongZhuFrazzoliAutom2016}. Section~\ref{sec:shots} draws connections to earlier works of singular filtering and introduces an accommodation to circumvent stability issues using the inner-outer factorization. It also contains the extension to time-varying systems via the Riccati equation. Section~\ref{sec:inverse} reinforces the connections to system inversion and concludes. The Appendix contains the proofs.

\section{Problem statement}\label{sec:problem}
SISE algorithms have been formulated for linear time-varying systems \cite{GillijnsDeMoorAutom2007a,GillijnsDeMoorAutom2007b,YongZhuFrazzoliAutom2016} and for nonlinear time-varying systems \cite{FangDeCallafonCortesAutom2013,KIM2020108588}. However for clarity in development, we consider the linear, time-invariant system with zero known control input,
\begin{align}
x_{t+1}&=Ax_t+Gd_t+w_t,\label{eq:state}\\
y_t&=Cx_t+Hd_t+v_t,\label{eq:output}
\end{align}
with $x_t\in\mathbb R^n,$ $d_t\in\mathbb R^m,$ $y_t\in\mathbb R^p.$ Zero-mean white noises $\{w_t\}$ and $\{v_t\}$ are independent and independent from $\{d_t\}$ and $x_0$. The covariance of $w_t$ is $Q\geq 0$ and the covariance of $v_t$ is $R>0.$ Denote the signal measurements $\mathbf Y^t\triangleq\{y_t,y_{t-1},\dots,y_0\}.$ The aim is to produce from $\mathbf Y^t,$ a recursive filtered state estimate, $\hat x_{t|t},$ and filtered and/or smoothed estimates, $\hat d_{t|t+1}$ or $\hat d_{t|t}$, depending on the properties of $G$ and $\begin{bmatrix}C&H\end{bmatrix}$. We make the following assumption.
\begin{assumption}\label{ass:crobs}
System (\ref{eq:state}-\ref{eq:output}) has $[A,C]$ observable, rank\,$G=m$, $[A,Q]$ reachable, and $R>0$.
\end{assumption}

Full-rank direct feedthrough, i.e. rank$(H)=m$, is treated in \cite{GillijnsDeMoorAutom2007b}; zero direct feedthrough, $H=0$, in \cite{GillijnsDeMoorAutom2007a} with rank$(CG)=m$; and, \cite{YongZhuFrazzoliAutom2016} provides a generalization, ULISE, with mixed rank properties between $H$ and $CG$.  A noise-free variant is treated in \cite{MarroZattoniAutom2010}. 
\section{Zero direct feedthrough}\label{sec:zdf}
For $H=0$ in \eqref{eq:output}, SISE from \cite{GillijnsDeMoorAutom2007a} is the recursion.
\begin{align}
X_t&=AP_{t-1}A^T+Q,\label{eq:xsise}\\
K_t&=X_tC^T(CX_tC^T+R)^{-1},\label{eq:sisek}\\
M_t&=[G^TC^T(CX_tC^T+R)^{-1}CG]^{-1}\nonumber\\
&\hskip 15mm\times G^TC^T(CX_tC^T+R)^{-1},\label{eq:sisem}\\
P_t&=(I-K_tC)\left[(I-GM_tC)X_t\right.\nonumber\\
&\hskip 8mm\left.\times(I-GM_tC)^T+GM_tRM_t^TG^T\right]\nonumber\\
&\hskip 18mm+K_tRM_t^TG^T,\label{eq:sisep}\\
\hat d_{t-1|t}&=M_t(y_t-CA\hat x_{t-1|t-1}),\label{eq:sised}\\
\hat x_{t|t}&=A\hat x_{t-1|t-1}+G\hat d_{t-1|t}+K_t\nonumber\\
&\hskip 8mm\times(y_t-CA\hat x_{t-1|t-1}-CG\hat d_{t-1|t}).\label{eq:sisex}\\
\cov(x_t\vert&\mathbf Y^t)=P_t,\label{eq:sisecovx}
\end{align}
under the following structural condition.
\begin{assumption}\label{ass:crank}
\begin{align}\label{eq:crank}
\text{rank}\,CG=m.
\end{align}
\end{assumption}
An immediate observation is that SISE contains no specific information related to a model for the unmeasured disturbance $d_t.$ Indeed, it is frequently claimed that signal $\{d_t:  t=0,1,\dots\}$ possesses no model whatsoever. Although, for bounded covariance $X_t$, i.e. when the algorithm is stable, the authors derived this version of SISE in \cite{BitmeadHovdAbooshahab:Autom2019} as a Kalman filter with $\{d_t\}$ modeled as a white noise process of unbounded variance. We shall return to this point later. Evidently, Assumption~\ref{ass:crank} requires $p\geq m$ and rank\,$C\geq$\,rank\,$G=m$. Firstly, we treat the square case, $p=m$, where the number of measurements equals the dimension of the disturbance input. Then we shall derive more general results.

\subsection{Square zero-feedthrough case}
From Assumption~\ref{ass:crank} when $p=m$, $CG$ is invertible. Since, from \eqref{eq:sisem}, $M_tCG=I$ or $M_t=(CG)^{-1}$, we have
\begin{align}
\hat d_{t-1|t}&=(CG)^{-1}(y_t-CA\hat x_{t-1|t-1}),\label{eq:dcg}\\
0&=y_t-CA\hat x_{t-1|t-1}-CG\hat d_{t-1|t},\label{eq:xinnov}\\
\hat x_{t|t}&=A\hat x_{t-1|t-1}+G\hat d_{t-1|t},\label{eq:dep}\\
&=[I-G(CG)^{-1}C]A\hat x_{t-1|t-1}+G(CG)^{-1}y_t.\label{eq:xcg}
\end{align}

This estimation algorithm: 
\begin{itemize}
\item[--] is time-invariant; 
\item[--] does not depend on $Q$ or $R$, the noise variances; 
\item[--] is independent from the covariance calculations.,
\item[--] has zero $\hat x_{t|t}$ innovations \eqref{eq:xinnov}, \eqref{eq:sisex}.
\end{itemize}
SISE reduces to (\ref{eq:dcg}-\ref{eq:xcg}).
\begin{align*}
\hat x_{t|t}&=[I-G(CG)^{-1}C]A\hat x_{t-1|t-1}+G(CG)^{-1}y_t,\\
\hat d_{t-1|t}&=-(CG)^{-1}CA\hat x_{t-1|t-1}+(CG)^{-1}y_t.
\end{align*}
Note that, using the matrix inversion lemma, we may rewrite the SISE $y_t$-to-$\hat d_{t-1|t}$ transfer function as
\begin{align}
\MoveEqLeft (CG)^{-1}\nonumber\\
\MoveEqLeft \hskip 4mm-(CG)^{-1}CA(zI-A+G(CG)^{-1}CA)^{-1}G(CG)^{-1}\nonumber\\
&=\left[CG+CA(zI-A)^{-1}G\right]^{-1},\nonumber\\
&=\left[zC(zI-A)^{-1}G\right]^{-1}.\label{eq:sqinv}
\end{align}
The filtered state estimate error satisfies
\begin{align*}
\tilde x_{t|t}&\triangleq x_t-\hat x_{t|t},\\
%&=Ax_{t-1}+Gd_{t-1}+w_{t-1}-A\hat x_{t-1|t-1}\\
%&\hskip 5mm-G(CG)^{-1}\left[C(Ax_{t-1}+Gd_{t-1}+w_{t-1})\right.\\
%&\hskip 10mm \left.+v_t-CA\hat x_{t-1|t-1|}\right],\\
&=[I-G(CG)^{-1}C]A\tilde x_{t-1|t-1}\\
&\hskip 5mm+[I-G(CG)^{-1}C]w_{t-1}-G(CG)^{-1}v_t.
\end{align*}
The stability of SISE, i.e. the boundedness of the covariance of $\tilde x_{t|t},$ depends on the eigenvalues of $[I-G(CG)^{-1}C]A$.

\begin{theorem}\label{thm:squarezero}
For system (\ref{eq:state}-\ref{eq:output}) with $p=m$ and subject to Assumption~\ref{ass:crank}, the eigenvalues of the SISE estimator system matrix, $\left[I-G(CG)^{-1}C\right]A,$ lie at the transmission zeros of the square transfer function $zC(zI-A)^{-1}G.$ Accordingly, the SISE estimator is asymptotically stable if and only if these transmission zeros all lie inside the unit circle.
\end{theorem}
The proof of this theorem follows immediately from \eqref{eq:sqinv}. An alternate is given in the Appendix for completeness and to establish connections to singular filtering. We note that condition \eqref{eq:crank} in Assumption~\ref{ass:crank} implies that $zC(zI-A)^{-1}G$ possesses exactly $n$ finite transmission zeros with exactly $m$ at zero.

We see that, in the square case, the poles of SISE can be located precisely at the transmission zeros of the $d_t$-to-$y_t$ transfer function. SISE therefore is performing system inversion to recover $\hat d_{t-1|t}$ from $\mathbf Y^t$. The dependent recursion \eqref{eq:dep} for $\hat x_{t|t}$ is a simulation of the state equation \eqref{eq:state} driven by $\hat d_{t-1|t}$. Effectively all the information in $\mathbf Y^t$ is used in generating the disturbance estimate, leaving simulation \eqref{eq:dep} to generate the state estimate.

When SISE is stable, it was shown in \cite{BitmeadHovdAbooshahab:Autom2019} that the state estimation algorithm implements a Kalman filter with a model for $\{d_t\}$ as a white noise of unbounded variance, $D$. In this case, the state estimation problem has driving noise variance $Q+GDG^T$ and measurement noise variance $R$. The identical filter, but not the covariances, will be achieved by taking driving noise $GDG^T$ for finite $D$ and $R\to 0$. That is, SISE is a singular filter. The connection to \cite{KJ:12} in the proof is to the equivalent result in Loop Transfer Recovery for LQG control. When one selects $R=0$, as opposed to $R\to 0$ from above, then the poles are placed at the transmission zeros. The limiting operation, on the other hand places the poles at the stable transmission zeros and the inverses of the unstable transmission zeros \cite{ShakedTAC1985}.

\subsection{Non-square zero-feedthrough case}\label{subsec:svd}
From Assumption~\ref{ass:crank}, we take $p\geq m$ and make a transformation of the output signal as follows. This is a variation on the technique of \cite{YongZhuFrazzoliAutom2016}. Take the singular value decomposition of $p\times m$ $CG$.
\begin{align*}
\texttt{svd}(CG)&=U\Sigma V^T,\\
&=\begin{bmatrix}U_m&U_{p-m}\end{bmatrix}\begin{bmatrix}\Sigma\\0\end{bmatrix} V^T.
\end{align*}
Define the $p\times p$ transformation
\begin{align}\label{eq:tdef}
\mathcal T&=\begin{bmatrix}U_m^T-U_m^TRU_{p-m}(U_{p-m}^TRU_{p-m})^{-1}U_{p-m}^T\\U_{p-m}^T\end{bmatrix},
\end{align}
and transform the original output signal, call it $\bar y_t$, 
\begin{align}\label{eq:trans}
y_t&=\mathcal T\bar y_t=\begin{bmatrix}C_1\\C_2\end{bmatrix}x_t+\begin{bmatrix}v_{1,t}\\v_{2,t}\end{bmatrix},
\end{align}
yielding
\begin{align*}
%\text{columns of }C_1&\in\text{range}\,G^T,\\
%\text{columns of }C_2&\in\text{null}\,G^T,\\
\det\,C_1G\neq 0,\;\;\;\;
C_2G=0,\;\;\;\;
\text{cov}\begin{bmatrix}v_{1,t}\\v_{2,t}\end{bmatrix}=\begin{bmatrix}R_1&0\\0&R_2\end{bmatrix}.
\end{align*}
\begin{theorem}\label{thm:nonsquarezero}
For system (\ref{eq:state}-\ref{eq:output}) with $p\geq m$ and subject to Assumptions~\ref{ass:crobs} and \ref{ass:crank}, if and only if the pair $[A(I-G(C_1G)^{-1}C_1), C_2]$ is detectable then the filtered state covariance, $P_t$, is bounded and converges to a limit $P_\infty$ as $t\to\infty.$

The corresponding gain matrices, $K_\infty$ and $M_\infty$, yield the limiting SISE system matrix, $(I-K_\infty C)(I-GM_\infty C)A,$ with all its eigenvalues strictly inside the unit circle.
\end{theorem}
The proof of this result appears in the Appendix and is based on proving that the state covariance satisfies a Riccati Difference Equation. Although this condition is not strictly the same as the condition in \cite{YongZhuFrazzoliAutom2016}, the theorem condition implies theirs. Hence, their condition is also necessary. Theorem~\ref{thm:nonsquarezero} similarly extends the condition in \cite{FangShiYiIJACSP2011}. The sufficient stability result in \cite{KIM2020108588} is predicated on $P_t$ being bounded a priori. We already know from Theorem~\ref{thm:squarezero} the eigenvalues of $A(I-G(C_1G)^{-1}C_1)$ are stable if and only if $zC(zI-A)^{-1}G$ is minimum-phase.

We see that, when $p>m$, the surfeit of measurements beyond those strictly needed to produce $\hat d_{t-1|t}$ are brought to bear on estimating $x_t$. The stability of SISE depends on either the square case yielding stability via Theorem~\ref{thm:squarezero}, i.e. via stable transmission zeros, or there being sufficient information in the additional measurements to stabilize the estimator.

When SISE is stable, then the algorithm implements a singular filter, as explained above. However, now the corresponding singular filter is \textit{partially singular}, a term introduced in \cite{PrielShakedCDC1986}. That is, the process noise variance is finite but the measurement noise variance, $R,$ is less than full rank rather than zero. The approach of \cite{PrielShakedCDC1986}, under the banner of stable optimal filtering, in this case involves precisely a succession of a singular estimator and followed by a regular estimator, as in SISE. The result in \cite{BitmeadHovdAbooshahab:Autom2019} derives this stable (partially) singular filter when the plant satisfies the conditions of Theorem~\ref{thm:nonsquarezero}.

\section{Nonzero direct feedthrough}\label{sec:ndf}
When $H\neq 0$ in \eqref{eq:output}, SISE alters. Gillijns and De~Moor \cite{GillijnsDeMoorAutom2007b} provide a SISE algorithm, subject to the following, for the time-invariant case.
\begin{assumption}\label{ass:hrank}
Rank\,$H=m$.
\end{assumption}
Subject to this assumption, the SISE formulation for time-invariant system (\ref{eq:state}-\ref{eq:output}) is
\begin{align}
\hat x_{t|t-1}&=A\hat x_{t-1|t-1}+G\hat d_{t-1|t-1},\label{eq:zxhat}\\
P^x_{t|t}&=\begin{bmatrix}A&G\end{bmatrix}\begin{bmatrix}P^x_{t-1|t-1}&P^{xd}_{t-1|t-1}\\P^{dx}_{t-1|t-1}&P^d_{t-1|t-1}\end{bmatrix}
\begin{bmatrix}A^T\\G^T\end{bmatrix}+Q,\nonumber\\
\tilde R_t&=CP^x_{t|t-1}C^T+R,\nonumber\\
M_t&=(H^T\tilde R_tH)^{-1}H^T\tilde R_t^{-1},\nonumber\\
\hat d_{t|t}&=M_t(y_t-C\hat x_{t|t-1}),\label{eq:zdhat}\\
P^d_{t|t}&=(H^T\tilde R_tH)^{-1},\nonumber\\
K_t&=P^x_{k|k-1}C^T\tilde R_t^{-1},\nonumber\\
\hat x_{t|t}&=\hat x_{t|t-1}+K_t(y_t-C\hat x_{t|t-1}-H\hat d_{t|t}),\label{eq:zfilt}\\
P^x_{t|t}&=P^x_{t|t-1}-K_t(\tilde R_t-HP^dH^T)K_t^T,\nonumber\\
P^{xd}_{t|t}&=\left(P_{t|t}^{dx}\right)^T=-K_tHP^d_{t|t}.\nonumber
\end{align}
When rank\,$H<m,$ \cite{YongZhuFrazzoliAutom2016} provide ULISE, a carefully developed SISE algorithm which uses the singular value decomposition as in Subsection~\ref{subsec:svd} but more widely to handle the more complicated interaction between filtered and smoothed estimates for $d_t$.

\subsection{Square full-rank case}
As with the $H=0$ case, we consider first rank\,$H=m$ and $m=p.$ That is $H$ is invertible and, since $M_tH=I$, $M_t=H^{-1}$. Then SISE reduces to the recursion
\begin{align}
\hat d_{t|t}&=H^{-1}(y_t-C\hat x_{t|t-1})\nonumber\\
0&=y_t-C\hat x_{t|t-1}-H\hat d_{t|t},\label{eq:zinno}\\
\hat x_{t+1|t}%&=\hat x_{t|t-1},\\
&=A\hat x_{t-1|t-1}+G\hat d_{t-1|t-1},\label{eq:dsimo}\\
&=(A-GH^{-1}C)\hat x_{t|t-1}+GH^{-1}y_t,\nonumber\\
\tilde x_{t+1|t}&=(A-GH^{-1}C)\tilde x_{t|t-1}+w_t-GH^{-1}v_t.\nonumber
\end{align}
\begin{theorem}\label{thm:squarenonzero}
For system (\ref{eq:state}-\ref{eq:output}) subject to Assumption~\ref{ass:hrank}, the eigenvalues of the SISE estimator system matrix, $A-GH^{-1}C,$ lie at the transmission zeros of the square transfer function $H+C(zI-A)^{-1}G.$ Accordingly, the SISE estimator is asymptotically stable if and only if these transmission zeros all lie inside the unit circle.
\end{theorem}
\textit{\underline{Proof}:}Applying the matrix inversion lemma to the square transfer function between $d_t$ and $y_t$,
\begin{align*}
&\left[H+C(zI-A)^{-1}G\right]^{-1}\\
&\hskip 10mm=H^{-1}-H^{-1}C(zI-A+GH^{-1}C)^{-1}GH^{-1}.
\end{align*}
The poles of the square direct feedthrough SISE lie at the transmission zeros of the $d_t$ to $y_t$ transfer function.\hfill$\Box$

Again, this result adds necessity to that of \cite{YongZhuFrazzoliAutom2016} in this case. Further, the result does not rely on optimality arguments. As in the square zero feedthrough case, the SISE estimator is time-invariant and independent from $Q$ and $R$, and the  state estimate filter innovations is zero. The condition rank\,$H=m$ ensures that all $n$ transmission zeros are finite. We note again that the state innovations sequence \eqref{eq:zinno} is zero and the filter \eqref{eq:dsimo} simulates $\hat x_{t+1|t}$ from $\hat d_{t|t}.$

\subsection{Non-square full-rank case}
The careful derivation of ULISE to accommodate rank\,$H\leq m$ is a central contribution of \cite{YongZhuFrazzoliAutom2016} and involves separation into subspaces. Take the singular value decomposition of $p\times m$ $H$ possessing rank $r$.
\begin{align*}
\texttt{svd}(H)&=U\Sigma V^T,\\
&=\begin{bmatrix}U_r&U_{p-r}\end{bmatrix}\begin{bmatrix}\bar H&0\\0&0\end{bmatrix} V^T.
\end{align*}
Matrices take on the $(r,p-r)$ structure.
\begin{align*}
H&=\begin{bmatrix}\bar H&0\\0&0\end{bmatrix},\;\;\;C=\begin{bmatrix}C_1\\C_2\end{bmatrix},\;\;G=\begin{bmatrix}G_1&G_2\end{bmatrix},\\
K_t&=\begin{bmatrix}K_{1,t}&K_{2,t}\end{bmatrix},\;\;\;M_t=\begin{bmatrix}M_{1,t}&M_{2,t}\end{bmatrix}.
\end{align*}
As earlier in \eqref{eq:tdef} and \eqref{eq:trans}, define the $p\times p$ transformation
\begin{align*}%\label{eq:tdef}
\mathcal T&=\begin{bmatrix}U_r^T-U_r^TRU_{p-r}(U_{p-r}^TRU_{p-r})^{-1}U_{p-r}^T\\U_{p-r}^T\end{bmatrix},
\end{align*}
and transform the original output signal, call it $\bar y_t$, 
\begin{align*}%\label{eq:trans}
y_t&=\mathcal T\bar y_t=\begin{bmatrix}\bar C_1\\\bar C_2\end{bmatrix}x_t+\begin{bmatrix}\bar H \\ 0\end{bmatrix}d_t+\begin{bmatrix}\bar v_{1,t}\\\bar v_{2,t}\end{bmatrix}.
\end{align*}
yielding $\det\,\bar H\neq 0$ and $\text{cov}\begin{bmatrix}\bar v_{1,t}\\\bar v_{2,t}\end{bmatrix}=\begin{bmatrix}\bar R_1&0\\0&\bar R_2\end{bmatrix}.$
%Without loss of generality, since $\mathcal T$ is a linear transformation of the output, take $H=\bar H.$ Then 
When rank\,$H=m$, $\bar H$ is $m\times m$ and we have the following result stemming from  $M_{1,t}\bar H=I_m$.
\begin{theorem}\label{thm:nonsquareFullH}
Subject to Assumptions~\ref{ass:crobs} and \ref{ass:hrank}, $p\geq m,$ SISE with feedthrough is stable if only if $[A-G\bar H^{-1}\bar C_1,\bar C_2]$ is detectable.
\end{theorem}
The proof of this result is in the Appendix. This extends the detectability condition\footnote{Note that \cite{YongZhuFrazzoliAutom2016} uses $p$ to denote our $m$.} of Theorem~5 of \cite{YongZhuFrazzoliAutom2016} to a necessary and sufficient condition for stability of SISE in this case. It also is the analog of Theorem~\ref{thm:nonsquarezero} for the full-rank feedthrough case.

%and divide $K_t$ and $M_t$ conformably: 
%$K_t=\begin{bmatrix}K_{1,t}&K_{2,t}\end{bmatrix},$ $M_t=\begin{bmatrix}M_{1,t}&M_{2,t}\end{bmatrix}.$
%Then the SISE state prediction recursion becomes
%\begin{align*}
%\hat x_{t+1|t}&=A\hat x_{t|t-1}+[AK_t+(G-AK_tH)M_t](y_t-C\hat x_{t|t-1}).
%\end{align*}
%Now, since $M_tH=I,$ $M_{1,t}=\bar H^{-1},$ and
%\begin{align*}
%\MoveEqLeft AK_t+(G-AK_tH)M_t\\
%&=A\begin{bmatrix}K_{1,t}&K_{2,t}\end{bmatrix}+G\begin{bmatrix}\bar H^{-1}&M_{2,t}\end{bmatrix}\\
%&\hskip 20mm-AK_{1,t}\bar H\begin{bmatrix}\bar H^{-1}&M_{2,t}\end{bmatrix},\\
%&=\begin{bmatrix}G\bar H^{-1}&AK_{2,t}+GM_{2,t}-AK_{1,t}\bar HM_{2,t}\end{bmatrix}.
%\end{align*}
An alternative way to view necessity is to write the system matrix of SISE as
\begin{align*}
\MoveEqLeft A-[AK_t+(G-AK_tH)M_t]C\\
&=A-G\bar H^{-1}\bar C_1\\
&\hskip 10mm-(AK_{2,t}+GM_{2,,t}-AK_{1,t}\bar HM_{2,t})\bar C_2.
\end{align*}
For this matrix to be stable, a multiple of $\bar C_2$ must stabilize $A-G\bar H^{-1}\bar C_1.$

\subsection{Less than full rank feedthrough}
We build again on the decomposition above of \cite{YongZhuFrazzoliAutom2016} and make the following assumption.
\begin{assumption}\label{ass:chrank}
$\text{rank}\,\bar C_2G_2=m-\text{rank}\,\bar H.$
\end{assumption}
In \cite{YongZhuFrazzoliAutom2016}, the authors derive a sufficient condition for stability which we now extend to necessity. 
\begin{theorem}\label{thm:nonsquarenonzeronon}
Subject to Assumptions~\ref{ass:crobs} and \ref{ass:chrank}, general feedthrough SISE is stable if only if $[A-G_1\bar H^{-1}\bar C_1,\bar C_2]$ is detectable.
\end{theorem}
This detectability condition is shown in \cite{YongZhuFrazzoliAutom2016} to be sufficient for stability by using the filter recursion for $\hat x_{t|t}$. If one calculates the alternative recursive prediction, $\hat x_{t|t-1},$ then it is evident that the ULISE system matrix is again of the form
\begin{align*}
\MoveEqLeft (A-G_1\bar H^{-1}\bar C_1)(I-\tilde L_t\bar C_2)(I-G_2M_{2,t}\bar C_2)\\
&=A-G_1\bar H^{-1}\bar C_1+W_t\bar C_2,
\end{align*}
for appropriate $W_t$. Evidently, this can be stable only if the detectability condition holds. 

\section{Upshots}\label{sec:shots}
\subsection{Stability and singular filtering}
The preceding analysis provides necessary and sufficient conditions for the stability of linear SISE algorithms. Further, for the square cases, it yields the precise locations on the algorithm poles and demonstrates that the emphasis is on $d_t$-to-$y_t$ system inversion to recover $d_t$ followed by best efforts to estimate the state. We have pointed out the successive estimation nature of SISE, as have others. The question remains as to actions to be taken when SISE proves to be unstable, noting these central properties:
\begin{enumerate}[label=(\roman*)]
\item SISE is stable when the $d_t$-to-$y_t$ system is stably invertible.
\item When SISE is stable, it corresponds (at least in the zero feedthrough case) to a singular Kalman filter.
\item Subject to: detectability of $[A,C]$, stabilizability of $[A,Q^\frac{1}{2}]$, and $R\stackrel +\to 0$; the singular Kalman filter is a stable estimator by construction. 
\item The stability conditions for the singular Kalman filter are more relaxed than Assumptions~\ref{ass:crank}, \ref{ass:hrank} or \ref{ass:chrank}.
\item When SISE proves to be unstable, it differs from the Kalman filter.
\end{enumerate}

\subsection{Kalman filtering for input and state estimation}
Denote the transfer function from $d_t$ to $y_t$ by $T(z)$. As derived in earlier sections, when $zT(z)$ has all its transmission zeros inside the unit circle, then SISE is guaranteed stable and is equivalent to a specific stable singular Kalman filter. Compute a discrete-time inner-outer factorization\footnote{Strictly speaking, this is a co-inner-co-outer factorization because of the ordering of $T_i$ and $T_o$ \cite{FrancisBook1987}. It can be obtained from the inner-outer factorization of $T^T$.}  \cite{GreenInnerOuterSCL1988,IonescuOaraTAC1996}.
\begin{align*}
T(z)&=T_o(z)T_i(z),
\end{align*}
with $m\times m$ $T_i(z)$ an inner function, i.e. stable and all-pass, and $p\times m$ $T_o(z)$ an outer function, i.e. all transmission zeros inside the unit circle.

\begin{figure}[h]
\begin{centering}
\includegraphics[width=33mm]{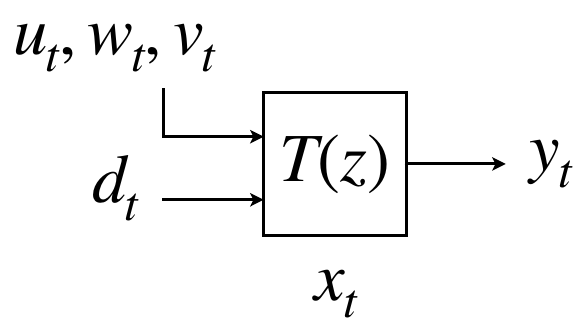}
\includegraphics[width=50mm]{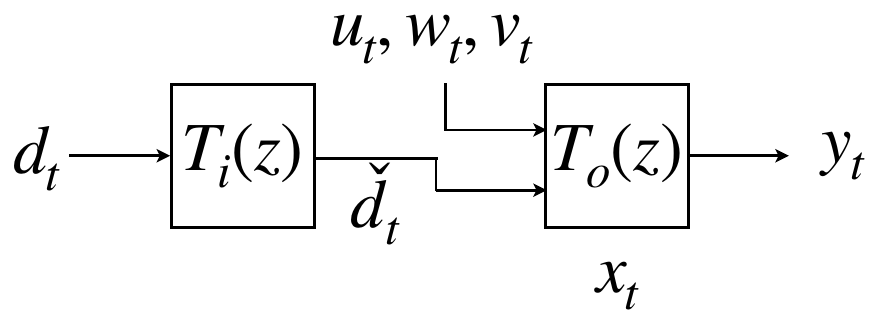}
\caption{System inner-outer factorization\label{fig:inOut}}
\end{centering}
\end{figure}

We note the following from the construction of the inner-outer factors.
\begin{lemma}
If $T(z)$ has realization 
\begin{align*}
x_{t+1}&=Ax_t+Bu_t+Gd_t+w_t,\\
y_t&=Cx_t+Du_t+Hd_t+v_t,
\end{align*}
then $T_o(z)$ has realization
\begin{align*}
x_{t+1}&=Ax_t+Bu_t+\check G\check d_t+w_t,\\
y_t&=Cx_t+Du_t+\check H\check d_t+v_t,
\end{align*}
where $\check d_t$ is the output of $T_i(z).$
\end{lemma}
That is, for the same initial conditions, $T_o(z)$ and $T(z)$ have the same states and have realizations which differ only in the $G$ and $H$ matrices.
This factorization is depicted in Figure~\ref{fig:inOut}.

Applying SISE or the singular Kalman filter for the outer function $T_o(z)$ to the signal $\{y_t\}$ yields estimates of $\{\check d_t\}$ and $\{x_t\}$ via a stable algorithm with guaranteed statistical and optimality properties. This follows since $T_o$ is stably invertible by construction. Further, this stability depends on the standard assumptions above for Kalman filter stability. If one uses SISE, then depending on the delay properties of $T_o(z)$, i.e. its behavior as $z\to\infty$, a modified variation of the algorithm and Assumption~\ref{ass:crank}, \ref{ass:hrank}, or \ref{ass:chrank} might be needed to accommodate $d_t$-to-$y_t$ invertibility. This is discussed further in Section~\ref{sec:inverse}.

To recover estimates for original system inputs $\{d_t\}$ from those for $\{\check d_t\}$  requires deconvolution (input estimation) without state estimation for the maximum-phase but stable system $T_i$. If delay is not an issue, then this can proceed stably via a fixed-interval smoother or  reverse-time input estimation.

If the state estimates of $x_t$ themselves are the objective, then the reconstruction of $\check d_t$ versus $d_t$ is immaterial. This is the nature of the problem addressed in partially-known power system state estimation \cite{AliMortenBobCCTA2019}.

The singular filters derived by Shaked and co-authors \cite{ShakedTAC1985,PrielShakedCDC1986,ShakedSorokaTAC1987} rely on the Return Difference Equality and spectral factorization for their calculation. In the case where the transmission zeros are unstable, the filter solution replaces them by their inverses akin to the inner-outer factorization.

The Kalman filter of \cite{BitmeadHovdAbooshahab:Autom2019} for $T_o(z)$ may be derived from the state-space model below with appropriate covariances,
\begin{align*}
\begin{bmatrix}x_{t+1}\\\check d_{t+1}\end{bmatrix}&=\begin{bmatrix}A&\check G\\0&0\end{bmatrix}\begin{bmatrix}x_t\\\check d_t\end{bmatrix}
+\begin{bmatrix}I_n&0\\0&I_m\end{bmatrix}\begin{bmatrix}w_t\\\delta_t\end{bmatrix},\\
y_t&=\begin{bmatrix}C&\check H\end{bmatrix}\begin{bmatrix}x_t\\\check d_t\end{bmatrix}+v_t,
\end{align*}
or using the direct construction as in \cite{BitmeadHovdAbooshahab:Autom2019}, which avoids an explicit model for $d_t$ but yields the same filter.

Marro and Zattoni \cite{MarroZattoniAutom2010} provide guidance on the recovery of the disturbance input signal when the $d_t$-to-$y_t$ system is non-minimum-phase. Their approach involves the approximate inversion of this system using a long delay to accommodate the nominal instability of this inverse. Such techniques are reminiscent of those advanced in \cite{WidrowWallach1994}. While the approach in \cite{MarroZattoniAutom2010} centers on state-estimation first, their development is geometric and noise free and so, it is unclear how this affects performance. Of course, the geometric analysis throws up the same initial reliance on minimum-phase zeros for stability and exact inversion.

\subsection{Extension to time-varying systems}
Developments so far have been limited to the time-invariant case and have availed themselves of concepts of transmission zeros, stable invertibility and inner-outer factorization, each of which is problematic to extend to time-varying systems. However, since alternative results have been phrased for the time-varying case, we consider this extension now, relying on examination of SISE recursions via Riccati difference equations in the proofs of Theorems~\ref{thm:nonsquarezero} and \ref{thm:nonsquareFullH}. 

Appealing to \cite{GillijnsDeMoorAutom2007a,GillijnsDeMoorAutom2007b} for the time-varying SISE algorithms in the case of Theorem~\ref{thm:nonsquarezero} and zero direct feedthrough, Riccati equation \eqref{eq:rdeZeroH} becomes
\begin{align*}
X_{t+1}&=\bar{A}_tX_t\bar{A}_t^T-\bar{A}_tX_tC_{2,t}^T(C_{2,t}X_tC_{2,t}^T+R_{2,t})^{-1}\\
&\hskip 34mm \times (\bar{A}_tX_tC_{2,t}^T)^T+\bar{Q}_t, 
\end{align*}
where,
\begin{align*}
\bar{A}_t&=A_t(I-G_{t-1}(C_{1,t}G_{t-1})^{-1}C_{1,t}),\\ \bar{Q}_t&=A_tG_{t-1}(C_{1,t}G_{t-1})^{-1}R_{1,t}(G_{t-1}(C_{1,t}G_{t-1})^{-1})^TA_t^T\\ &\hskip 8mm +Q_t,
\end{align*}
and, in the case of full-rank feedthrough, \eqref{eq:rde2} becomes
\begin{align*}
X_{t+1}&=\hat A_tX_{t}\hat A_t^T - (\hat A_tX_{t}\bar C_{2,t}^T)(\bar C_{2,t}X_{t}\bar C_{2,t}+\bar R_{2,t})^{-1}\\ 
& \hskip 34mm\times  (\hat A_tX_{t}\bar C_{2,t}^T)^T+\hat Q_t,
\end{align*}
where,
\begin{align*}
\hat A_t= A_t-G_t\bar H_t^{-1}\bar C_{1,t}, \;\;\; \hat Q_t =Q_t+G_t\bar H_t^{-1}\bar R_{1,t}\bar H_t^{-T}G_t^T.
\end{align*}
with now time-varying quantities $\{A_t,G_t,\dots,\}$. We may appeal to standard sufficient results, e.g. \cite{Jazwinski:70,Anderson&Moore:79} and Theorem 5.3 in \cite{AndersonMooreDetectabilitySIAM1981}, on the exponential stability of the Kalman filter subject to uniform reachability and detectability. Subject to the uniform satisfaction of time-varying equivalents of Assumptions~\ref{ass:crobs}, \ref{ass:crank} and/or \ref{ass:hrank} as appropriate, this extends these stability conditions to the uniformly time-varying case.

\section{System Inversion and SISE}\label{sec:inverse}
For the square cases of SISE satisfying Assumptions~1 or 2, we were able to demonstrate that the SISE $d_t$-estimator implements exactly the left inverse of the $y_t$-to-$d_t$ system. The simultaneous $x_t$-estimate is the state of the inverse system the stability of which depends on the transmission zeros of the original system. %Inverse systems have been studied for some time \cite{SainMasseyTAC1969,MoylanTAC1977,SundaramURL}.

Conditions for left invertibility of a linear time-invariant system are provided by Sain and Massey \cite{SainMasseyTAC1969} and for stable invertibility by Moylan \cite{MoylanTAC1977} via the Rosenbrock system matrix. Both papers construct the inverse system. Moreover, in \cite{SainMasseyTAC1969,SundaramURL}, left invertibility with delay, $L$, is studied, where stacked measurements $\begin{bmatrix}y_t^T&y_{t+1}^T&\dots&y_{t+L}^T\end{bmatrix}^T$ are used to estimate $d_t$. Marro and Zattoni blend into this picture stable approximate inversion with delay.

From \cite{SainMasseyTAC1969}, we see that, for any $p\geq m$:
\begin{itemize}
\item[--] Assumption~\ref{ass:crank} is the left invertibility condition for $C(zI-A)^{-1}G$ with delay one. 
\item[--] Assumption~\ref{ass:hrank} is the left invertibility condition for $H+C(zI-A)^{-1}G$ with delay zero.
\item[--] Assumption~\ref{ass:chrank} is the left invertibility condition for $H+C(zI-A)^{-1}G$ with delay one.
\end{itemize}
Given the input recovery objective of SISE, this is not surprising. But it is interesting to tie these ideas more closely.

It is worth remarking that many presentations of SISE algorithms make connections to `unbiasedness' and `optimality' of the state estimate. As \cite{MarroZattoniAutom2010,BitmeadHovdAbooshahab:Autom2019} demonstrate, the probabilistic concept of unbiasedness is really tied to a geometric property of the algorithms and the nature of certain subspaces. The optimality of the state estimates is within the class of estimators already satisfying the geometric constraints. As is evident from, say, Theorems~\ref{thm:squarezero} and  \ref{thm:nonsquarezero} and the Riccati equation proof, there is no degree of freedom left for the state estimator in the square case and limited degrees of freedom in the non-square case. Indeed, in the square cases, (\ref{eq:dep}-\ref{eq:dsimo}) show that $\hat x_{t|t}$ is computed by system simulation using the estimated input; the measurements play no further part. The detectability conditions on Theorems~\ref{thm:nonsquarezero} and \ref{thm:nonsquareFullH} show how the remaining degrees of freedom are used in the Riccati difference equations \eqref{eq:rde1} and \eqref{eq:rde2}.

In conclusion, the paper attempts to unify the collection of SISE algorithms by revealing their explicit connections to system inversion to recover the otherwise unmodeled disturbance input $d_t$ followed by their `best efforts' subsequent estimation of the state $x_t$. The result has been to develop necessary and sufficient conditions for stability, at least in the linear time-invariant case, in terms of the transmission zeros of the $d_t$-to-$y_t$ plant and then the detectability of the subsequent state estimator.

\section{Appendix}
\subsection*{Proof of Theorem~\ref{thm:squarezero}}
We loosely follow a calculation from Maciejowski \cite{KJ:12}. From \eqref{eq:xcg} and the system equations the (time-invariant) transfer function from $d_t$ to $\hat x_{t|t}$ via $y_t$ is given by
\begin{align}
\Psi(z)&=\left\{zI-\left[I-G(CG)^{-1}C\right]A\right\}^{-1}zG(CG)^{-1}\nonumber\\
&\hskip 30mm \times C(zI-A)^{-1}G,\label{eq:zbit}\\
&=\left\{zI-\left[I-\Pi\right]A\right\}^{-1}z\Pi(zI-A)^{-1}G,\nonumber
\end{align}
where we have used $\Pi\triangleq G(CG)^{-1}C$. Write
\begin{align*}
\MoveEqLeft \left[zI-(I-\Pi)A\right]^{-1}z\Pi\\
&=\left[zI-(I-\Pi)A\right]^{-1}\left[z\Pi-zI+(I-\Pi)A\right]+I,\\
&=-\left[zI-(I-\Pi)A\right]^{-1}(I-\Pi)(zI-A)+I.
\end{align*}
Then, since $(I-\Pi)G=0,$
\begin{align}
\MoveEqLeft\Psi(z)=\nonumber\\&-\left[zI-(I-\Pi)A\right]^{-1}(I-\Pi)(zI-A)(zI-A)^{-1}G\nonumber\\
&\hskip 30mm+(zI-A)^{-1}G,\nonumber\\
&=-\left[zI-(I-\Pi)A\right]^{-1}(I-\Pi)G+(zI-A)^{-1}G,\nonumber\\
&=(zI-A)^{-1}G.\label{eq:zbit2}
\end{align}
From \eqref{eq:zbit}, $\Psi(z)$ is the product of two transfer functions and nominally should have $2n$ poles; those at the eigenvalues of $A$ and those at the eigenvalues of $(I-\Pi)A$. The transfer function $zC(zI-A)^{-1}G$ has McMillan degree $n$ with $n$ finite transmission zeros. We see from \eqref{eq:zbit2} that only poles at the eigenvalues of $A$ are present in $\Psi$. This implies that the poles due to the eigenvalues of $(I-\Pi)A$ cancel the transmission zeros of $zC(zI-A)^{-1}G.$

\subsection*{Proof of Theorem~\ref{thm:nonsquarezero}}
Define the following quantities.
\begin{align*}
\mathcal{H}_t &=  GM_t-K_tCGM_t+K_t,\;\;\;Z=C_1G,\\
Y&=(CX_tC^T+R)^{-1}=\begin{bmatrix}Y_1&Y_2\\Y_2^T&Y_3\end{bmatrix},
\end{align*}
where $Y$ is divided conformably with $C$ and $v_t$ in \eqref{eq:trans}.

From \eqref{eq:sised} and \eqref{eq:sisex} the filtered prediction error satisfies
\begin{align*}
\tilde{x}_t &\triangleq x_t-x_{t|t}\\
&=(I-\mathcal{H}_tC)A\tilde{x}_{t-1} + (I-\mathcal{H}_tC)w_{t-1} -\mathcal{H}_tv_{t}.
\end{align*}
Whence,
\begin{align*}
P_{t|t}&= \operatorname{cov}(x_t|\mathbf{Y_t})\\
&= (I-\mathcal{H}_tC)(AP_{t-1|t-1}A^T+Q) (I-\mathcal{H}_tC)^T\\ &\hskip 8mm+\mathcal{H}_tR\mathcal{H}_t^T.
\end{align*}
Using \eqref{eq:xsise} yields
\begin{align}
X_{t+1} &= A\left((I-\mathcal{H}_tC)X_t(I-\mathcal{H}_tC)^T+\mathcal{H}_tR\mathcal{H}_t^T\right)A^T\nonumber\\ &\hskip 50mm+Q. \label{eq:Xt}
\end{align}
We show that this discrete Lyapunov equation is also a Riccati difference equation by substituting for $\mathcal H_t$ using $CG=\begin{bmatrix}Z^T&0\end{bmatrix}^T$.

\begin{align}
\mathcal{H}_t &= G\left(\begin{bmatrix}Z^T&0\end{bmatrix}Y\begin{bmatrix}Z\\0\end{bmatrix}\right)^{-1}\begin{bmatrix}Z^T&0\end{bmatrix}Y-X_t\begin{bmatrix}C_1^T&C_2^T\end{bmatrix}\nonumber \\
&\hskip 8mm Y\begin{bmatrix}Z\\0\end{bmatrix}\left(\begin{bmatrix}Z^T&0\end{bmatrix}Y\begin{bmatrix}Z\\0\end{bmatrix}\right)^{-1}\begin{bmatrix}Z^T&0\end{bmatrix}Y&\nonumber\\ 
&\hskip 12mm+X_t\begin{bmatrix}C_1^T&C_2^T\end{bmatrix}Y\nonumber\\
&=\begin{bmatrix}GZ^{-1}&GZ^{-1}Y_1^{-1}Y_2\end{bmatrix}+X_t\begin{bmatrix}0&C_2(Y_3-Y_2^TY_1^{-1}Y_2)\end{bmatrix} \label{eq:H},
\end{align}
Using partitioned matrix inversion with $Y$ gives
\begin{align*}
(Y_3-Y_2^TY_1^{-1}Y_2)&=(C_2X_tC_2^T+R_2)^{-1} \\
Y_1^{-1}Y_2&=-(C_1X_tC_2^T)(C_2X_tC_2^T+R_2)^{-1} .
\end{align*}
Substituting this into \eqref{eq:H} and \eqref{eq:Xt} gives the following Riccati difference equation.
\begin{align}
X_{t+1}&=\bar{A}X_t\bar{A}^T-(\bar{A}X_tC_2^T)(C_2X_tC_2^T+R_2)^{-1}\nonumber\\
&\hskip 34mm \times (\bar{A}X_tC_2^T)^T+\bar{Q}, \label{eq:rdeZeroH}
\end{align}
where,
\begin{align}
\bar{A}&=A(I-G(C_1G)^{-1}C_1),\nonumber\\ \bar{Q}&=AG(C_1G)^{-1}R_1(G(C_1G)^{-1})^TA^T+Q.\label{eq:rde1}
\end{align}
Appealing to Theorem~14.3.1 \cite{KailathSayedHassibi2000} (p.~510), provided $[\bar A,\bar Q^\frac{1}{2}]$ is stabilizable and $[\bar A, C_2]$ is detectable, then $X_t$ converges to the maximal solution of the algebraic Riccati equation, which is stabilizing.

Now, since by assumption $[A,Q^\frac{1}{2}]$ is stabilizable, there exists a $\mathcal K$ such that $A-Q^\frac{1}{2}\mathcal K$ is stable. Taking,
\begin{align*}
\bar Q^\frac{1}{2}&=\begin{bmatrix}Q^\frac{1}{2}&AG(C_1G)^{-1}R^\frac{1}{2}\end{bmatrix},
\end{align*}
and $\bar{\mathcal K}=\begin{bmatrix}\mathcal K^T&R^\frac{T}{2}\end{bmatrix}^T$, $\bar A-\bar Q^\frac{1}{2}\bar{\mathcal K}$ also is stable. So stabilizability of $[A,Q^\frac{1}{2}]$ implies stabilizability of $[\bar A, \bar Q^\frac{1}{2}]$.

%we know that $X_t$ converges to the solution of a corresponding discrete-time algebraic Riccati equation (DARE) if $X_0 \geq 0$. Given the assumption that $(A,Q^{1/2})$ is stabilizable, the corresponding DARE has a stablizing positive definite solution if and only if $(\bar{A},C_2)$ is detectible as can be seen from theorem E.5.1 of  [\cite{KailathSayedHassibi2000}, p.~783]. This proves theorem \ref{thm:nonsquarezero}.\\
%Note: our assumption of stabilizability of $(A,Q^{1/2})$ is sufficient for the stability condition in theorem E.5.1 of  \cite{KailathSayedHassibi2000} which is $(F^s, Q^{s/2})$ being stabilizable, where $F^s$ and $Q^{s/2}$ are given in [\cite{KailathSayedHassibi2000}, p.~774]. This can be shown as follows.  $(F^s, Q^{s/2})$ is stabilizable if and only if there exists a matrix $N=\begin{bmatrix}N_1&N_2\end{bmatrix}$ such that $F^s+NQ^{s/2}$ is stable. Choosing $N$ such that $N_1$ stabilizes $A+N_1Q^{1/2}$ and $F^s+NQ^{s/2} = A+N_1Q^{1/2}$ can be done, showing the sufficiency of our condition, namely, the stabilizability of $(A,Q^{1/2})$, for applying theorem E.5.1 of \cite{KailathSayedHassibi2000}.

\subsection*{Proof of Theorem~\ref{thm:nonsquareFullH}}
The proof parallels that of Theorem~\ref{thm:nonsquarezero}. Substitute \eqref{eq:zdhat} and \eqref{eq:zfilt} into \eqref{eq:zxhat} to yield
\begin{align*}
\hat{x}_{t+1|t}&=(A-\mathcal L_tC)\hat{x}_{t|t-1}+\mathcal L_ty_t,
\end{align*}
where $\mathcal L_t=AK_t-AK_tHM_t+GM_t$. Then
\begin{align*}
\tilde{x}_{t+1|t}&\triangleq x_t-\hat x_{t+1|t},\\
&=(A-\mathcal L_tC)\tilde{x}_{t|t-1}+w_t-\mathcal L_t v_t,\\
X_{t+1} &= (A-\mathcal L_tC)X_{t}(A-\mathcal L_tC)^T+\mathcal L_tR\mathcal L_t^T+Q,
\end{align*}
with $X_{t+1}\triangleq \operatorname{cov}(x_{t+1}|\mathbf{Y}^t)$. Dividing $K_t$ and $M_t$ conformably with $C^T$: 
$K_t=\begin{bmatrix}K_{1,t}&K_{2,t}\end{bmatrix},$ $M_t=\begin{bmatrix}M_{1,t}&M_{2,t}\end{bmatrix},$ one arrives directly at the following Riccati difference equation.
\begin{align}\label{eq:rde2}
X_{t+1}&=\hat AX_{t}\hat A^T - \hat AX_{t}\bar C_2^T(\bar C_2X_{t}\bar C_2+\bar R_2)^{-1}\bar C_2X_{t}\hat A^T\nonumber
\\ & \hskip 40mm +\hat Q,
\end{align}
where,
\begin{align*}
\hat A= A-G\bar H^{-1}\bar C_1, \;\;\; \hat Q =Q+G\bar H^{-1}\bar R_1\bar H^{-T}G^T.
\end{align*}
The proof follows as that for Theorem~\ref{thm:nonsquarezero} using \cite{KailathSayedHassibi2000}.

\bibliographystyle{unsrt}
\bibliography{/Users/bob/tex/bob}

\end{document}